\documentclass[slac_one]{revtex4}
\usepackage{graphicx}
\usepackage{fancyhdr}
\begin{document}

\title{Commissioning of the ATLAS Electron and Photon Trigger Selection} %% Paper title goes here


\author{Cibr\'{a}n Santamarina R\'{\i}os\footnote{for the ATLAS collaboration.}}
\affiliation{Department of Physics\\
                  McGill University\\
                 Montreal, QC, Canada\\
                 }

\begin{abstract}
Since the start-up of the LHC end of 2009, the trigger commissioning
is in full swing. The ATLAS trigger system is divided into three levels:
the hardware-based first level trigger, and the software-based second
level trigger and Event Filter, collectively referred to as the High
Level Trigger (HLT). Initially, events have been selected online based
on the Level-1 selections with the HLT algorithms run but not rejecting
any events. This has been an important step in the commissioning of
these triggers to ensure their correct functioning and subsequently to
enable the HLT selections. Due to increasing LHC luminosity and the
large QCD cross section, this has been a vital step to select leptons
from J/$\Psi$, bottom, charm, W and Z decays.

This presentation gives an overview of the trigger performance of the
electron and photon selection. 
Comparisons of the online selection variables with the offline 
reconstruction are shown as well as comparisons of data with MC 
simulations on which the current selection tuning is performed.
\end{abstract}

\maketitle

\thispagestyle{fancy}


\section{LHC and ATLAS}

The Large Hadron Collider (LHC) is the new proton-proton collider operating at the European Laboratory for Particle Physics (CERN). It is the highest energy particle accelerator ever built providing two high intensity beams of protons
accelerated at energies of 3.5 TeV (7 TeV in the future).
The LHC creates unique conditions for the study of new phenomena in Particle Physics
including the search for the Higgs Boson, the test of Supersymmetry and other Dark Matter models, etc.

ATLAS is a multipurpose experiment, at the LHC, whose detector is barrel structured. The ATLAS spectrometer is the largest ever built in particle physics. It is a complex technological system that
is capable of reconstructing tracks, measure the energies and identify the different kind of particles with the highest accuracy in the most intense rate of data conditions.

\section{The ATLAS Trigger System}

The ATLAS trigger is a system designed to identify, in real time,
potentially interesting interactions out of the billions produced per
second at the Large Hadron Collider (LHC).

The ATLAS trigger is tiered in
three levels that provide the necessary rejection to select
approximately 200~Hz of collision data from the LHC beam crossing rate
of 40~MHz.  Each trigger level has a different design and performance
characteristics dictated by the expected input event rates and available
processing time.
\begin{itemize}
\item The first level (L1), that only uses the information from the calorimeter and the muon detectors, uses dedicated hardware processors that run fast reconstruction
algorithms with coarse granularity and basic calibrations. The latency for its decision is 2.5~$\mu$s, reducing the rate to $\sim$75~kHz.
\item The second and third trigger levels (called L2 and EF, respectively) are both software based.
L2 runs a fast dedicated algorithm that uses the full detector granularity and reconstructs only a
small region of the detector, called Region of Interest or RoI, as found by L1. The average
processing time per event is $\sim 40$~ms and reduces the rate to about 3~kHz.
\item The EF executes offline like algorithms, that could run in RoI based reconstruction or full
event access. The average processing time per event at the EF is $\sim 4$~s and its output rate is 200 Hz.
\end{itemize}

\section{The ATLAS L1 e-$\gamma$ Selection}

\begin{figure}[htb]
\centering
\includegraphics[height=2.1in]{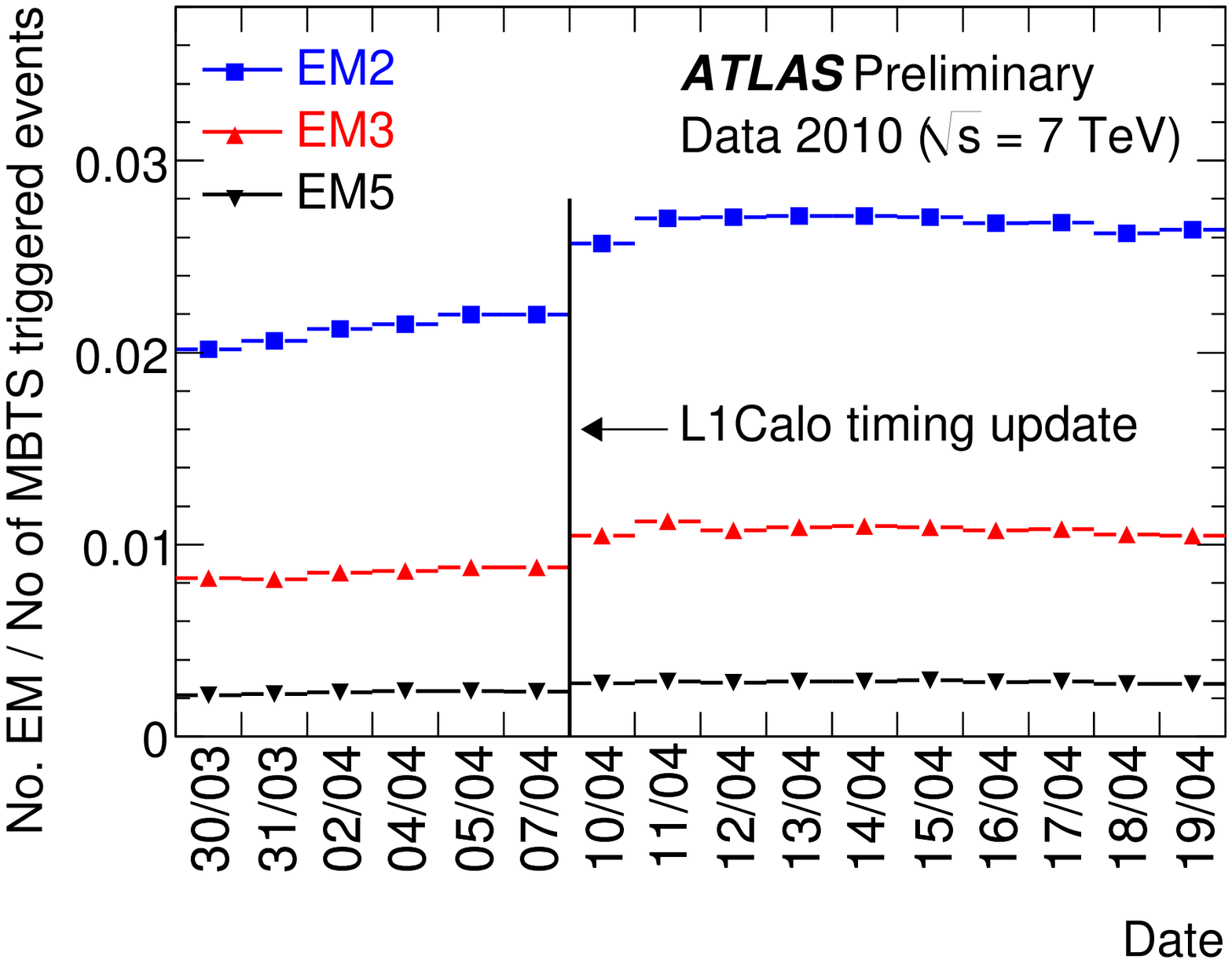}
\includegraphics[height=2.1in]{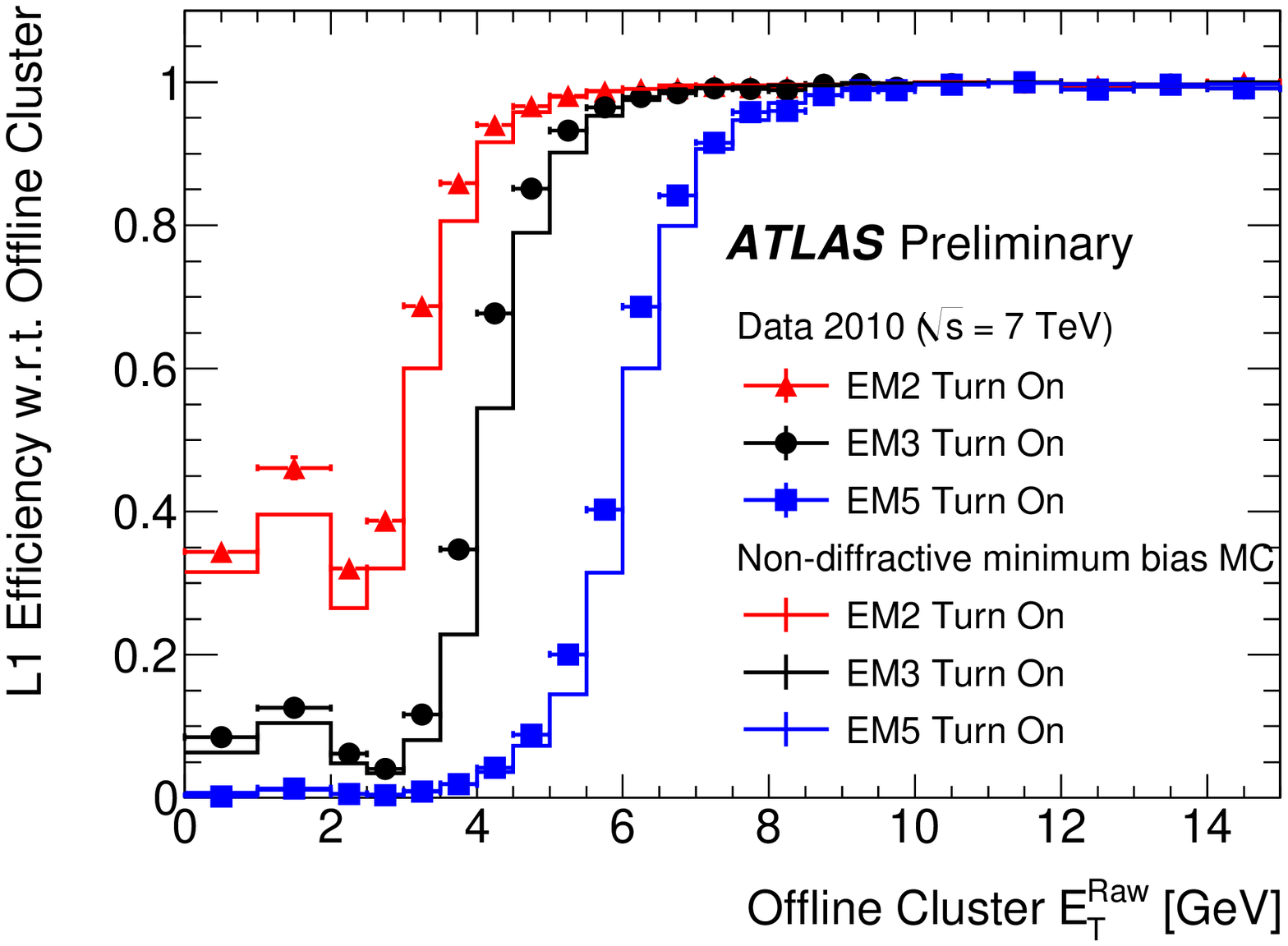}
\caption{{\bf Left:} {\em Percentage of MBTS Events fulfilling the L1 EM2, EM3 and EM5 thresholds.}
{\em Note the rate increase due to an increased efficiency after the trigger timing update after April the 10th.}
{\bf Right:} {\em Turn-on curves for the mentioned thresholds in data and MC samples.}}
\label{fig:l1res}
\end{figure}

The L1 electron-photon selection is purely calorimetric. The ATLAS
calorimeter is segmented into electromagnetic and hadronic trigger
towers which are analogue sums of calorimeter cell signals with a granularity of $0.1\times 0.1$
($\eta \times \phi$) and transmitted as analogue values.
A PreProcessor digitises, synchronises, calibrates and performs bunch-crossing identification.
The e-$\gamma$ and $\tau$ candidates are found in the Cluster Processing Modules (CPM).
Merger modules collect, sum and send results to the Central Trigger Proccessor, a module that makes the trigger decisions.
The algorithm implemented in the CPM for e-$\gamma$ reconstruction is a
$4\times 4$ Towers Sliding Window. The cluster candidate is required to satisfy
three conditions:
\begin{itemize}
\item Being a local $E_T$ (transverse energy) maximum above a configurable
threshold.
\item The total $E_T$ in the EM isolation region $<$ EM isolation threshold.
\item The total $E_T$ in the hadronic isolation region $<$  hadronic isolation threshold.
\end{itemize}
The commisioning of the e-$\gamma$ triggers is made by analyzing the real collision data taken
with a minimum bias trigger (MBTS) and studying the offline reconstruction performance of
events tagged with an EM trigger. The results are compared to a non-diffractive
minimum bias Monte Carlo sample proccessed through the full simulation
chain of the ATLAS detector and are summarized in Fig.~\ref{fig:l1res}.

\section{The ATLAS HLT e-$\gamma$ Selection}

\begin{figure}[htb]
\centering
\includegraphics[height=2.1in]{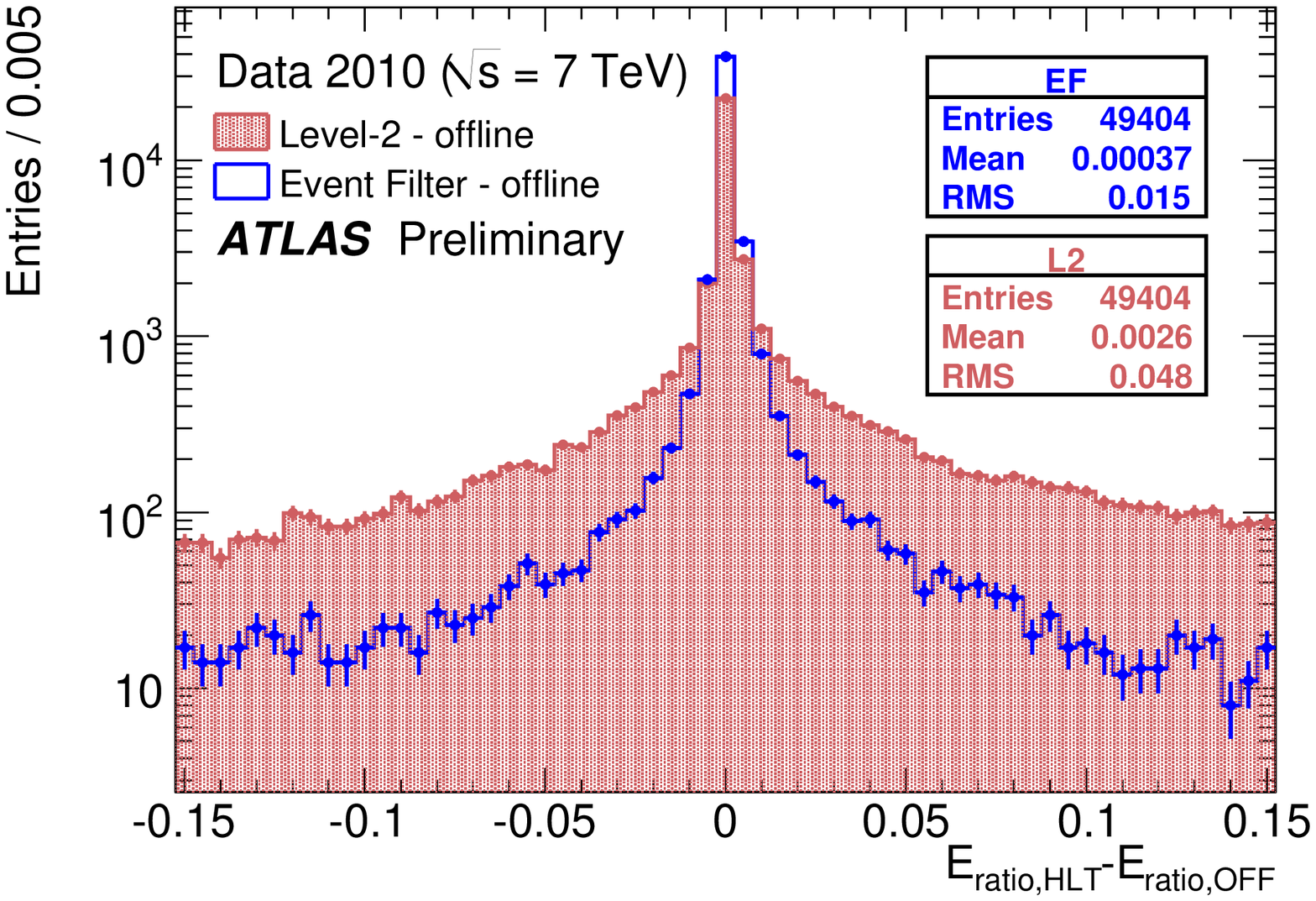}
\includegraphics[height=2.1in]{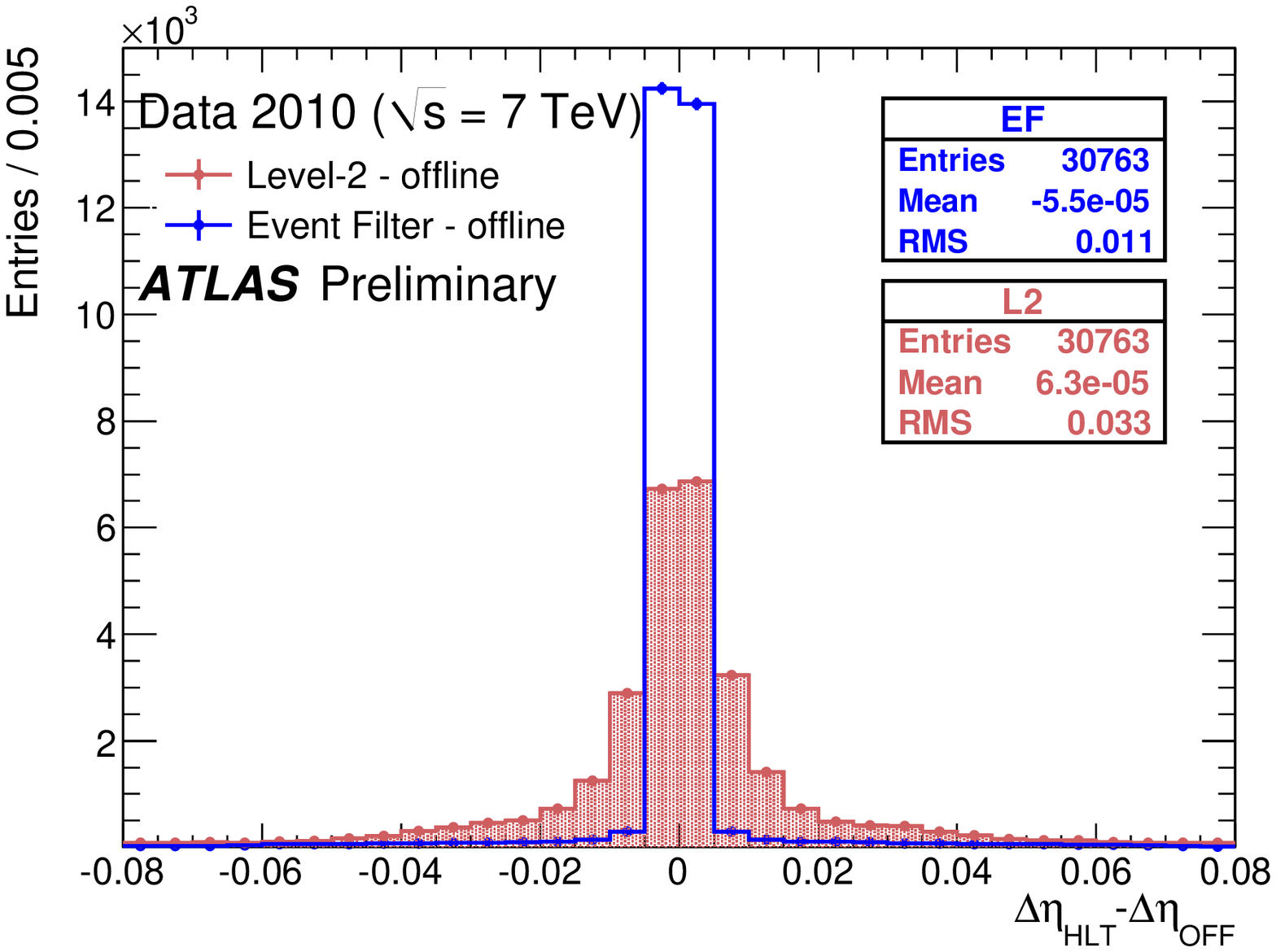}

\caption{{\bf Left:} $R_{\eta}$ {\em resolution with respect to the offline reconstruction.} {\bf Right:}{\em $\eta$ resolution of the tracking for the HLT.}
%{\bf Right:}{\em Track reconstruction efficiency as function of the transverse momentum.}}
}
\label{fig:magnet}
\end{figure}

The L2 combines Calorimeter and Inner Detector information. The full granularity of the
detector within a Region of Interest , as being defined by L1, is considered. At this level photon conversions and effects from electron Bremsstrahlung are not corrected.
At the EF an offline like reconstruction is applied. The most
refined recovery methods are included or could be included at this level. For instance, Photon conversions electron Bremsstrahlung corrections are not yet but could be considered. Both L2 and EF apply requirements on some cluster shape variables. In particular, in the leftmost plot of Fig.~\ref{fig:magnet}, the resolution of $R_{\eta}$, the ratio between the energy in a $3\times 7$ tower window and the energy in a $7\times 7$ window in EM Sampling 2, is shown.

%
%\begin{center}
%\resizebox*{0.6\columnwidth}{!}{\includegraphics{images/fruit/L2trig.eps}}
%\end{center}
%
For the electron identification a reconstructed track matching the calorimeter cluster is required. Again
L2 uses fast algorithms and the EF uses mostly offline algorithms.
The distance between the calorimeter
cluster and the reconstructed track as well as the ratio between the transverse momentum of the track
and the cluster $E_T$ are used as selection variables. The data performance matches very well the MC
prediction. The tracking performance is summarized in the rightmost plot of Fig.~\ref{fig:magnet}.

\section{Conclusions}

The ATLAS trigger for e-$\gamma$ selection shows an excelent performance with respect
to offline candidates, both in terms of efficiency and rejection power. The system has been running in stable mode  during most of the beam time at the LHC.

%%%%%%%%%%%%%%%%%%%%%%%%%%%%%%%%%%%%%%%%%%%%%%%%%%%%%%%%%%%%%%%%%%%%%%%%%
%%
%%   use this format to include a LaTeX table  into your paper
%%
%\begin{table}[t]
%\begin{center}
%\begin{tabular}{l|ccc}  
%Patient &  Initial level($\mu$g/cc) &  w. Magnet &  
%w. Magnet and Sound \\ \hline
% Guglielmo B.  &   0.12     &     0.10      &     0.001  \\
% Ferrando di N. &  0.15     &     0.11      &  $< 0.0005$ \\ \hline
%\end{tabular}
%\caption{Blood cyanide levels for the two patients.}
%\label{tab:blood}
%\end{center}
%\end{table}
%%%%%%%%%%%%%%%%%%%%%%%%%%%%%%%%%%%%%%%%%%%%%%%%%%%%%%%%%%%%%%%%%%%%%%%%%%%


\begin{thebibliography}{99}

%%
%%  bibliographic items can be constructed using the LaTeX format in SPIRES:
%%    see    http://www.slac.stanford.edu/spires/hep/latex.html
%%  SPIRES will also supply the CITATION line information; please include it.
%%

\bibitem{Mele07}
V. Dao,
{\em Commissioning of the ATLAS Electron and Photon Trigger Selection},
ATL-COM-DAQ-2010-097. To appear in CALOR proceedings.

\bibitem{CSCBook} ATLAS Collaboration, G Aad et al, {\em Expected Performance of the ATLAS Experiment - Detector, Trigger and Physics}, CERN-OPEN-2008-020 (2008).

\end{thebibliography}
\end{document}